# Surface properties of colloidal particles affect colloidal self-assembly in evaporating self-lubricating ternary droplets

Olga Koshkina,*,†,# Lijun Thayyil Raju,‡,# Anke Kaltbeitzel,† Andreas Riedinger,† Detlef Lohse,‡,¶ Xuehua Zhang,*,§,‡ Katharina Landfester*,†

†Max Planck Institute for Polymer Research, Ackermannweg 10, 55128, Mainz. Germany

‡Physics of Fluids Group, Max Planck Center for Complex Fluid Dynamics, MESA+ Institute and J. M. Burgers Center for Fluid Dynamics, University of Twente, PO Box 217, 7500 AE, Enschede, The Netherlands

¶Max Planck Institute for Dynamics and Self-Organisation, 37077 Göttingen, Am Fassberg 17, Germany

§Department of Chemical and Materials Engineering, University of Alberta, 12-380 Donadeo Innovation Centre for Engineering, Edmonton, T6G1H9 Alberta, Canada

# These authors contributed equally to the work

E-mail:

o.koshkina@utwente.nl, xuehua.zhang@ualberta.ca, landfester@mpip-mainz.mpg.de

**ABSTRACT**

In this work, we unravel the role of surface properties of colloidal particles on the formation of supraparticles (clusters of colloidal particles) in a colloidal Ouzo droplet. Self-lubricating colloidal Ouzo droplets are an efficient and simple approach to form supraparticles, overcoming the challenge of the coffee stain effect in situ. Supraparticles are an efficient route to high-performance materials in various fields, from catalysis to carriers for therapeutics. Yet, the role of the surface of colloidal particles in the formation of supraparticles using Ouzo droplets remains unknown. Therefore, we used silica particles as a model system and compared sterically stabilized versus electrostatically stabilized silica particles — positively and negatively charged. Additionally, we studied the effect of hydration. Hydrated negatively charged silica particles and sterically stabilized silica particles form supraparticles. Conversely, dehydrated negatively charged silica particles and positively charged amine-coated particles form flat film-like deposits. Notably, the assembly process is different for all the four types of particles. The surface modifications alter (a) the contact line motion of the Ouzo droplet and (b) the particle−oil and particle−substrate interactions. These alterations modify the particle accumulation at the various interfaces, which ultimately determines the shape of the final deposit. Thus, by modulating the surface properties of the colloidal particles, we can tune the shape of the final deposit, from a spheroidal supraparticle to a flat deposit. In the future, this approach can be used to tailor the supraparticles for applications such as optics and catalysis, where the shape affects the functionality.

## Introduction

The assembly of colloidal particles to supraparticles is an emerging strategy to produce functional materials.[1, 2] The supraparticles gain additional functionalities compared to the individual colloidal particles, as a result of the structural arrangement of the particles, combinations of different materials, and collective effects.[1] Hence, supraparticles find applications in a wide range of fields such as optics, catalysis, magnetic materials, and drug delivery.[3-10] Evaporation-induced self-assembly (EISA) of colloidal particles in self-lubricating ternary droplets, i.e. colloidal Ouzo droplets, is an efficient route to produce supraparticles.[11, 12] Although it is known that the surface properties of colloidal particles can influence their assembly in droplets,[13, 14] the role of the particles' surface on the formation of supraparticles in Ouzo droplets is still unknown. Here we show that the surface properties of the colloidal particles affect and can even impede the formation of supraparticles in colloidal 'Ouzo' droplets.

Ouzo droplets are named after the Greek aperitif Ouzo which contains water, ethanol, and oil, typically anise oil or *trans*-anethole.[15, 16] Colloidal Ouzo droplets are obtained by addition of colloidal particles to the Ouzo mixture.[11] During the evaporation of the Ouzo droplet, the concentration of ethanol decreases faster than that of water. As a result, the oil becomes less soluble, phase-separates, and forms a lubricating oil ring at the contact line.[15-19] This oil ring prevents pinning and leads to the formation of a supraparticle instead of the coffee ring.[11, 12]

The supraparticle formation can be influenced by parameters such as the size of the colloidal particles[12] or the oil-to-particle ratio[11] in the initial Ouzo mixture. However, what will happen when the surface properties of the particles are varied? Modifying the surface properties of the colloidal particles can alter the interactions of these particles with the various interfaces present in a sessile Ouzo droplet.

Depending on the surface properties, particles can adsorb onto an oil-water interface, stabilizing droplets of a dispersed phase in a continuous phase, forming Pickering emulsions.[20-22]

In a Pickering emulsion, the stability of the dispersed phase depends on the contact angle of the particle, and thus on the hydrophobicity. Colloidal particles with contact angle close to 90° provide the highest stability to the Pickering emulsions.[23] This stability arises from the high energy required to remove the particles that sit at the liquid-liquid interface with contact angle close to 90° [refer to SI Section S1 for more details]. Furthermore, it is known from the studies on non-evaporating Ouzo mixtures that when the system is in the metastable Ouzo-regime, the oil droplets are generally negatively charged and stabilized by electrostatic repulsion.[24, 25] Recently, it was shown that in the Ouzo regime, the addition of nanoparticles results in oil droplets with a narrow size-distribution.[26] Therefore, both hydrophobicity and the electrostatic interactions between the particles and oil can affect the strength of particle-oil interactions and hence the particle-assembly.

The surface properties of the colloidal particles can influence the arrangement of the particles during the evaporation of colloidal droplets even without oil. Particularly for colloidal droplets sitting on a substrate, the attractive particle-substrate interactions can suppress the coffee ring[14, 27, 28] and cause a disordered arrangement of particles on the substrate.[29] Moreover, modifying the hydrophobicity of particles can lead to the adsorption of particles at the liquid-air interface[28] or gelation close to the liquid-air interface,[30] leading to different deposition patterns than the classical coffee stain pattern.[28, 30-32] Specifically for supraparticles, the interactions between the particles affect the morphology, internal structure, dispersibility, and porosity of the final supraparticle.[13, 33-35] Thus, the surface properties of colloidal particles influence: (i) the interparticle interactions,[13, 33-37] (ii) the interactions between particles and substrate,[14, 27-29] and (iii) the behavior of the particles at the liquid-gas interface.[28, 30]

Therefore, to study the effect of surface properties on supraparticle formation in evaporating colloidal Ouzo droplets, we compared the following colloids: electrostatically stabilized negatively-charged silica particles, electrostatically stabilized positively charged (amine-coated) silica particles, and sterically-stabilized silica particles that were coated with a short poly(ethylene glycol) (PEG). Additionally, as hydration of the particle surface can influence the properties of colloidal particles, in particular the wetting properties,[38-41] we further compared hydrated and dehydrated silica particles. We observed that the surface modifications of the particles affect their interactions with the oil droplets and the substrate. The hydrated negatively charged silica particles and the PEGylated particles form spheroidal supraparticles. On the contrary, the negatively charged dehydrated particles as well as the positively charged amine-coated particles formed flat film-like deposits. Thus, the surface properties of particles play a crucial role in the formation of supraparticles in self-lubricating ternary droplets, and even suppress the supraparticle formation in some cases. This study provides a guideline to exploit the surface characteristics of the colloids for tuning the properties of supraparticles, particularly in the production of multi-component materials for applications such as drug delivery[3] and optoelectronics.[42, 43]

## Results and discussion

### Colloidal particles used to study the surface effects

To study the effect of surface modification of colloidal particles on the formation of supraparticles in self-lubricating droplets, we used different silica particles, as shown in Figure 1. Negatively-charged silica particles with a diameter of 450 nm were synthesized using the Stöber method, and labeled with rhodamine B-precursor during the polycondensation for fluorescence microscopy (Figure 1, Table 1). These silica particles were subsequently modified with amine groups or PEG, leading to positively charged or sterically stabilized PEGylated silica particles, respectively (Figure 1). The particles were dried after the synthesis and purification.[44] To study the effect of hydration in subsequent experiments, the particles were redispersed in water and incubated overnight, leading to *hydrated particles*. The *dehydrated particles* were obtained by resuspending the dried particles in absolute ethanol. Thus, all the particles display similar size but differ in their surface properties (Figure 1, Table 1).

Dynamic light scattering (DLS), transmission electron microscopy (TEM), and zeta potential measurements in aqueous dispersion showed the success of synthesis and modification (Figure 1, Table 1). All particles had a size of around 450 nm, based on TEM, and a narrow size distribution in the solvents present in the Ouzo-mixture, namely ethanol and water, as demonstrated by low polydispersity indexes (PDI) in DLS. After the PEGylation, the charge of silica particles increases from -58 to -41mV (Table 1). PEGylated particles still display negative charge, as we used a trifunctional silica precursor with a low molecular weight PEG for the surface modification (see Figure 1). Hence, some silanol groups are still present on the surface under the PEG-layer. As a result, the negative zeta potential can still be detected at the slip plane.[45] The amine-modified particles were positively charged (Table 1). After the modification with amine groups, the silica particles remain colloidally stable in ethanol as evident from the DLS results (Table 1). Slight agglomeration is observed upon the transfer into aqueous media due to non-covalent interactions between the particles; similar behavior was reported earlier in the literature.[44] Overall, the measurements (Table 1) show that the particles are colloidally stable in all solvents present in self-lubricating droplets.

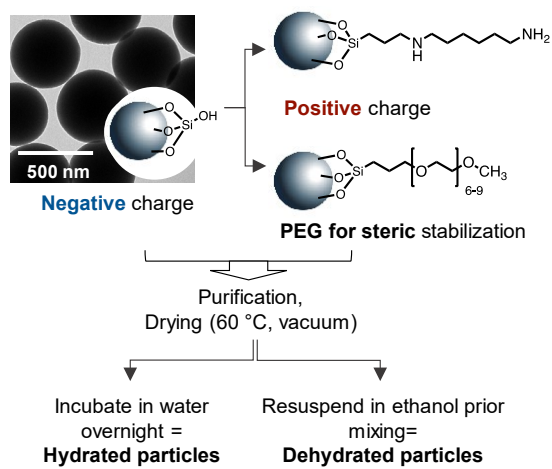

**Figure 1. Surface modification of silica particles.** Stöber silica particles were covalently modified with (6-aminohexyl)-aminopropyltrimethoxysilane or PEG-triethoxysilane, leading to positively-charged amine-modified or sterically-stabilized PEGylated particles respectively. After the synthesis, all particles were purified and dried at 60°C. To subsequently study the effect of hydration, particles were resuspended and incubated in water (hydrated particles). Alternatively, particles were resuspended in ethanol prior preparation of the Ouzo-mixtures (dehydrated particles). Transmission electron micrograph of negatively charged silica particles is shown. Scale bar 500 nm.

Table 1. Characterization of nanoparticles used for supraparticle formation[a].

| Particles | $D$ / nm (TEM) | Hydrated particles | | | Dehydrated particles | |
|---|---|---|---|---|---|---|
| | | $D_h$ / nm (DLS, 3 mM KCl in $H_2O$) | PDI | $\zeta$-pot. / mV (3 mM KCl in $H_2O$) | $D_h$ / nm (DLS in EtOH) | PDI |
| Unmodified silica | 449 ± 23 | 471 | 0.07 | -58 | 469 | 0.09 |
| PEGylated silica | 445 ±19 | 483 | 0.17 | -41 | 483 | 0.08 |
| Amine-coated silica | 443 ± 18 | 605 | 0.18 | 37 | 474 | 0.09 |

[a] $D$ is the diameter of the particles, measured using TEM. $D_h$ is the hydrodynamic diameter and PDI is the polydispersity index measured using DLS. $\zeta$-pot is the zeta potential of the particles.

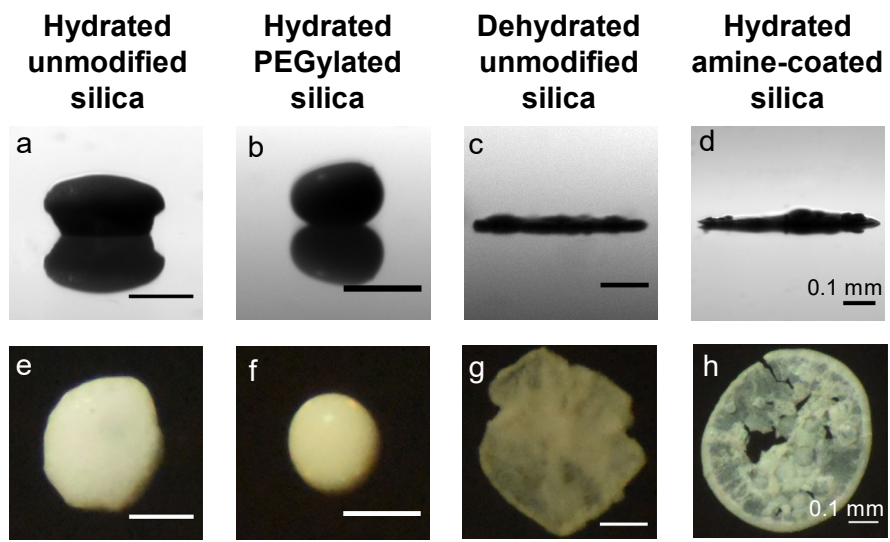

Figure 2. The surface modification of the colloidal particles affects the shapes of the final deposits obtained from evaporating Ouzo droplets. (a-d) Side view shadowgraphy images of the final deposit obtained after evaporation of droplets that were loaded with different silica particles, as mentioned at the top of the panel. (e-h) Top view images of the final deposits, corresponding to the side view images. Hydrated unmodified silica particles and PEGylated silica particles form

**supraparticles while the dehydrated unmodified silica particles and amine-coated silica particles form flat film-like deposits. Scale bar 0.1 mm.**

**Effect of the surface modification on the final particle deposit**

To study whether the surface modification of silica particles affects the particle assembly in an evaporating colloidal Ouzo droplet, we carried out evaporation experiments. The initial Ouzo mixture is a homogenous solution of water, ethanol, and *trans*-anethole. This mixture is in the one-phase region in the ternary phase diagram (see Methods), but also contains the colloidal particles. A droplet of ≈1 µL of this colloidal Ouzo was deposited on a hydrophobized glass surface. After all the liquids in the droplet have evaporated, a deposit of silica particles is left behind. The shape of the final deposits varies from a supraparticle to a flat deposit, as becomes evident in the side-view and the top-view images (Figure 2a-h; see Figure S2-S5 for images from repeated measurements). Thus, the surface properties of colloidal particles strongly affect the shape of the final deposit, as shown in Figure 2a-h.

Hydrated silica particles formed supraparticles that have a spheroidal shape, similar to the observation in our previous study[12] (side view and top view, Figure 2a and e respectively). PEGylated silica particles also formed spheroidal supraparticles (Figure 2b and f), but showed variations in the final shape (see SI Figure S4 for more images). Differently, the dehydrated unmodified silica particles and amine-coated particles led to flat film-like deposits (Figure 2 c, d, g, and h). The area-equivalent diameter and the aspect ratio further quantify these differences in the shape of the final deposit (Figure S5). The surface hydration did not affect the shape of the final deposits with amine-modified and PEGylated particles considerably. Therefore, throughout this paper, we only show the hydrated PEGylated and hydrated amine-coated silica particles in the figures of the main text; the corresponding images of dehydrated PEGylated and dehydrated amine-coated silica particles are shown in SI. Overall, there is a correlation between the surface modification of colloidal particles and the shape of the deposits obtained after the evaporation of colloidal Ouzo droplets (Figure 2).

**Effect of particle-surface modification on the evaporation dynamics**

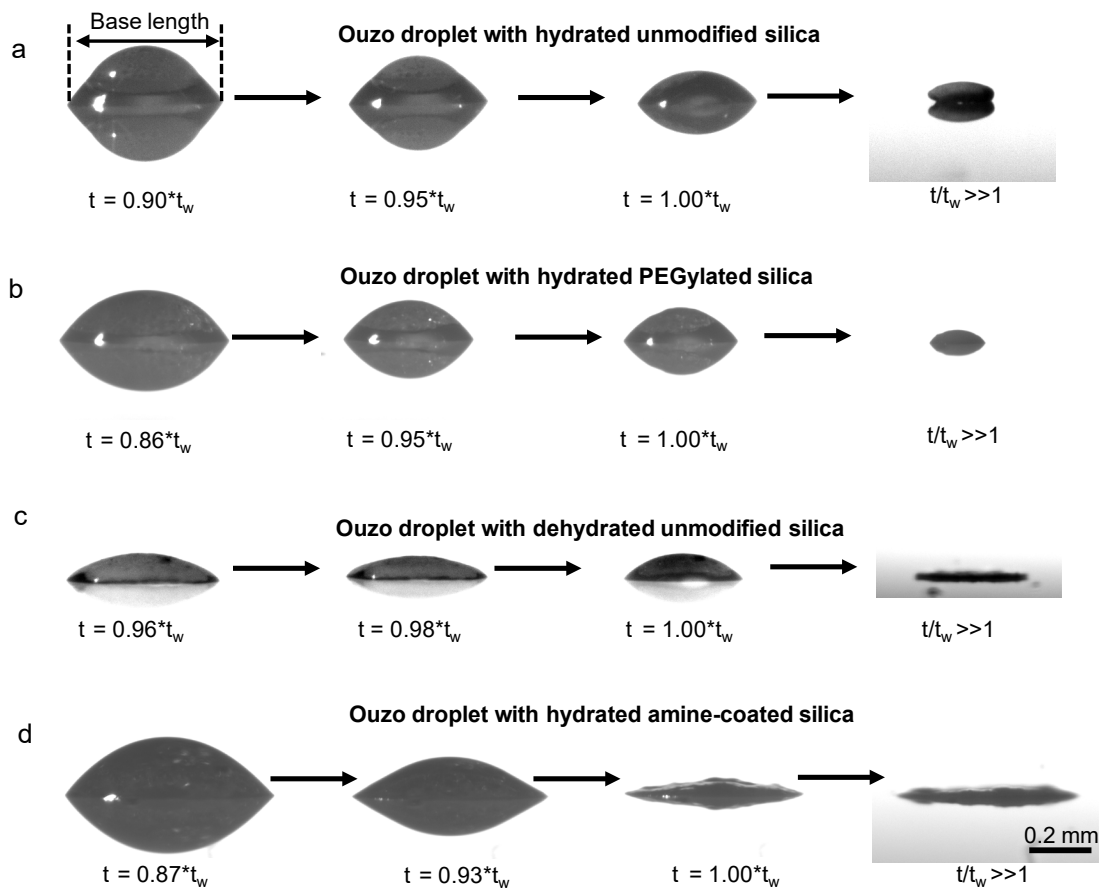

Figure 3. Side-view shadowgraph images of colloidal Ouzo droplets, at different time instances, showing the drop evaporation and steps leading to final deposit formation for (a) hydrated unmodified silica, (b) hydrated PEGylated silica, (c) dehydrated unmodified silica and (d) hydrated amine-coated silica. Hydrated unmodified silica particles and PEGylated silica form supraparticles, while dehydrated unmodified silica and amine-coated silica particles form flat deposits. Scale bar 0.2 mm.

To explain the differences in the shape of the final deposit obtained from evaporating colloidal Ouzo droplets, we observed the evaporation using top-view imaging and side-view shadowgraphy (Figure 3, Supplementary Video V1). From these results, we determined the volume and base diameter over the course of the evaporation, which are shown in Figure 4a and b (see SI, Figure S6-S8 for additional data, plots of height of the droplets over time, and plots of dehydrated PEGylated and dehydrated amine-coated particles). The time axis was normalized using $t_w$, which corresponds to the time when the plots of both volume vs time and base length vs time reach a plateau (Figure 4). At the time $t=t_w$, most of the water has evaporated, leaving the assembled silica particles and oil behind. It is very likely that some trace amount of water is present in the silica particles and oil even after $t=t_w$, which we cannot detect. The average value of $t_w$ is 930 ± 170 s, for the initial droplet volume of 0.7 ± 0.1 μL, as estimated from the side view measurements; the error shows the standard deviation of all measurements. The droplet volumes show similar trend over time for all the droplets (Figure 4a). Differently, the base length shows similar trend only in the initial but not in the final part of the evaporation.

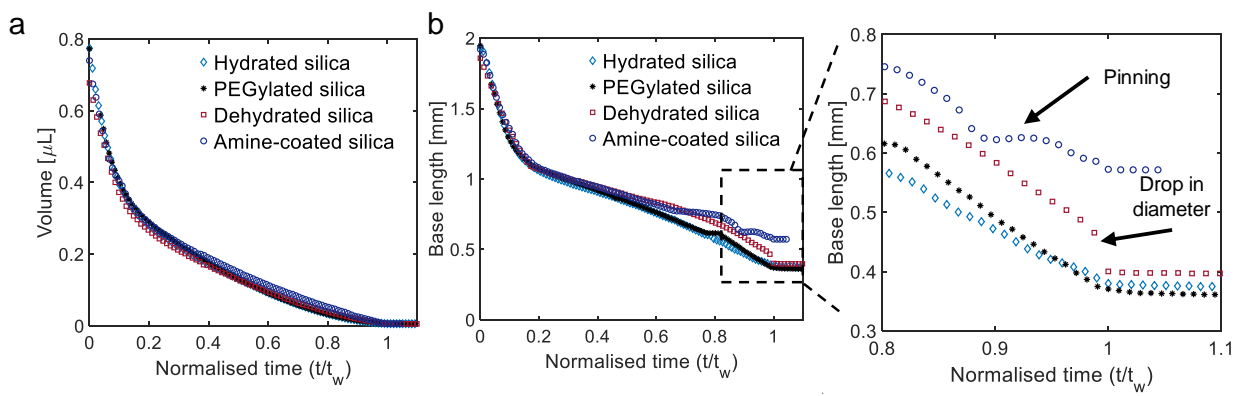

**Figure 4.** (a) Evolution of the droplet volume with time *t* normalized to time $t_w$. There is no appreciable difference in the volume evolution between droplets with different particles. (b) The base length shows different trends in the final phase of the evaporation process (zoomed region), depending on the type of silica particles. Refer to SI Section 3 for additional plots and repeat measurements. $t_w$ is estimated from the variation in volume and base length over time. $t_w$ corresponds to the time when the plots of both volume vs time and base length vs time reach a plateau. At $t=t_w$, most of the water has evaporated but some trace amount of water could be left behind, which cannot be measured.

In the final phase of evaporation, the base length reveals clear differences between colloidal droplets that formed supraparticles and flat film-like deposits (Figure 4b zoomed). Thus, the base length of the droplets with hydrated silica particles and PEGylated silica particles keeps on decreasing until $t/t_w = 1$, when most of the water has evaporated (Figure 4b and Figure 3a,b).

Conversely, for dried silica particles, the base length drops suddenly by 0.04 ± 0.02 mm, at $t/t_w$ = 0.98 ± 0.01 (Figure 4b and Figure 3c). This peculiar behavior is more evident in an increase in the height of the droplet by 14 ± 6 % (or 11 ± 3 µm) at $t/t_w$ = 0.98 ± 0.01 (Figure S6c). Finally, after the subsequent evaporation of the oil, a flat film-like deposit is obtained (Figure 3c). Differently, for droplets containing amine-coated particles the base length shows significant pinning and depinning (close to $t/t_w$ ≈ 0.7 and $t/t_w$ ≈ 0.9; Figure 4b). The final base length, at $t/t_w$=1, is ≈0.6 mm, which is slightly larger than in all other cases (≈0.4 mm, Figure 4b). Thus, the analysis of the side view images reveals differences in the motion of the contact-line of the droplet, leading to differences in the shape of the final deposit. This behavior indicates that the modifications in surface-properties of the particles affects the self-lubrication.

**Observing particle-oil interactions in static Ouzo mixtures**

Shadowgraphy raises the question whether the surface modifications affect the interaction of the particles with the oil in the droplet. Therefore, we first accessed whether the colloids interact with the oil in the Ouzo mixture and observed colloidal Ouzo-mixtures with fluorescence confocal microscopy under non-evaporating conditions. For these experiments, we chose a composition of the Ouzo mixture that is expected to correspond to the composition in the evaporating droplet close to the spinodal, where the two-phase region forms. This composition was estimated based on Diddens *et al*.[46] and the phase diagram by Sitnikova *et al*.[19] We confirmed that the composition was close to the spinodal by preparation of Ouzo-mixtures without colloidal particles (see methods for the exact composition). Confocal microscopy showed clear differences in the particle distribution around the oil microdroplets for the different colloidal particles (Figure 5).

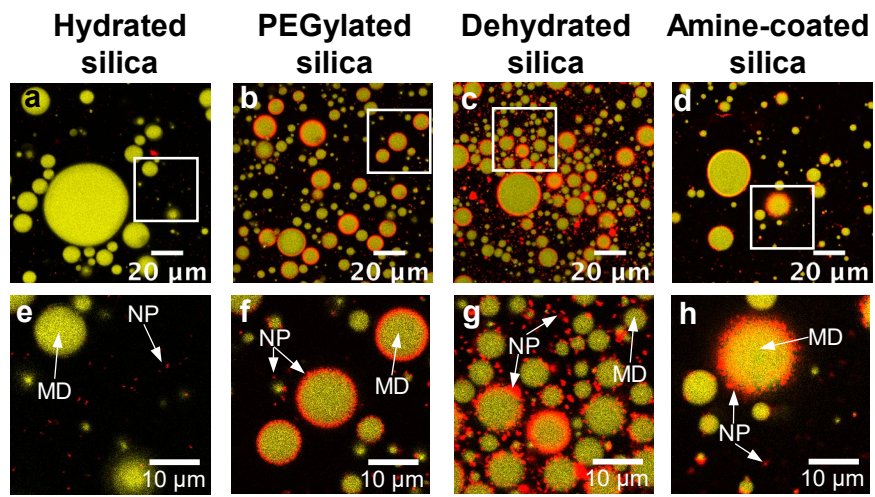

**Figure 5. Effects of surface modifications on particle-oil interactions.** The overlay of fluorescence signals from the particles (red) and oil (yellow) is shown. (a, e) Hydrated silica particles (NP) do not interact with oil microdroplets (MD). In contrast, PEGylation (b, f), dehydration (c, g), and surface modification with amine (d, h) lead to the adsorption of the particles (NP) onto the surface of oil microdroplets (MD). Thus, the surface modifications and dehydration lead to formation of structures similar to Pickering emulsions. The composition of the mixtures was the following (by *weight*): *t*-anethole / particles / water / ethanol: 0.020 / 0.0016 / 0.55 / 0.43. Overview images (a-d): Scale bar 20 µm. Zoomed in images (e-f): Scale bar 10 µm. See also Figures S9-S14 and figures in supplementary file Z1 for larger overview images, images for the Ouzo mixtures containing dehydrated amine-coated silica and dehydrated PEGylated silica, and for repetition of the experiments.

The surface modification with PEG and amine groups, as well as drying, altered the interactions between the colloidal particles and oil, leading to adsorption of the particles (red) onto the surface of the oil microdroplets (yellow, confocal microscopy Figure 5b-d,f-h). On the contrary, in the mixture with hydrated negatively charged particles, only polydisperse oil droplets were observed due to the spontaneous emulsification and phase separation (Figure 5a; see Table S1 for drop sizes). The particles remained dispersed in the continuous water-rich phase (Figure 5e, Figure S9). The automated analysis of the red florescence signal revealed high signal intensities in the red fluorescence channel around the oil microdroplets, confirming the differences between hydrated non-modified and other particles (see Table S1 and section S4 for further analysis of microscopy data). The particle adsorption increases the stability of the oil droplets, as also recently shown by Goubault *et al.*[26] in their comprehensive study on the stability and size-distribution of oil droplets in ouzo systems containing hydrophobic particles. In our experiments, the particle coated oil droplets remain stable for at least one week (see Supplementary Section

S4.2 and Figure S12-S14 for stability and size distribution of the phase separated oil-droplets immediately after preparation and after one week).

The particle coated oil droplets in Figure 5 appear similar to Pickering emulsions. Hence, we call these particle-coated oil droplets "Pickering microdroplets" in the following. Note that there is a difference between the formation of conventional Pickering emulsions and the Pickering microdroplets. The droplets in Pickering emulsions are obtained under energy input, using homogenization.[21] In contrast, in Ouzo mixtures, the oil phase-separates spontaneously upon an increase in the concentration of a non-solvent (water) in the mixture, forming oil droplets that get coated with the colloidal particles.

The differences in interactions between the oil droplets and the differently modified colloidal particles are likely caused by the changes in hydrophobicity of particles and electrostatic effects. Hydrated silica particles are highly hydrophilic and typically do not adsorb onto the oil droplets, as observed in classical Pickering emulsions.[21] PEG is soluble in water and in a broad range of organic solvents, which makes PEG suitable to stabilize particles in aqueous media.[47-49] Additionally, PEG can be used as a solvent for *t*-anethole suggesting attractive interactions between PEG and *t*-anethole.[47] Based on this affinity of PEG towards both water and *t*-anethole, it can be expected that PEGylated particles can adsorb onto the surface of the oil microdroplets.

For dehydrated silica particles, the removal of the hydrate water layer can lead to a slight increase in hydrophobicity, and consequently, to the adsorption of particles onto the oil microdroplets.[26, 38] It is known from literature that by suspending the particles in short-chain alcohols, the surface of particles can be made more hydrophobic due to the physical adsorption of the alcohol molecules.[21] Furthermore, it was recently shown that certain oils, depending on their solubility in water, can increase the hydrophobicity of the hydrophilic particles in-situ.[50] As a result, the hydrophilic particles can still form Pickering emulsions.[50] Thus, similar to these studies, the adsorption of ethanol and *t*-anethole on the silica particles could further enhance the hydrophobicity of the particles, enabling them to adsorb onto the oil microdroplets. In the case of amine-coated particles, the amine precursor that contains an alkyl-chain (Figure 1) can lead to an increase in hydrophobicity, as indicated by a slightly better dispersibility of amine-coated particle in ethanol compared with water (Table 1). Furthermore, the surface of oil droplet is usually negatively charged in an Ouzo mixture.[24,25] Hence, the electrostatic interactions between positively-charged particle and negatively charged interface of the oil droplets can further affect the adsorption. Overall, both the dehydration and the chemical modifications change the interactions between the oil droplets and the particles, and the stability of the oil droplets in the non-evaporating emulsions.

**Particle assembly observed through confocal microscopy**

As the surface modifications affect the particle-oil interactions, we carried out fluorescence confocal microscopy of the evaporating droplets to further understand the formation of different deposits. Confocal microscopy of the evaporating droplets confirms that modification of the particle surface alters the oil droplets that are nucleated on the substrate (Figure 6). We further compared the time sequence images obtained from an overlay of the fluorescence emission of silica particles (rhodamine labeled, red) and oil (perylene labeled, yellow) in Figure 7 and Figure 8. Bottom view refers to the images at a horizontal plane close to the substrate, and side view refers to the vertical cross sections that are reconstructed from stacks of horizontal planes (Figure 7 and Figure 8). These images show the spatial distribution of oil (yellow) and the silica particles (red) over time, as well as the formation of the final deposit. We summarize all the results from confocal microscopy of evaporating droplets as a scheme in Figure 9. In the following, we first discuss the formation of the final deposit for each type of particles separately, and later compare the mechanisms of the deposit formation.

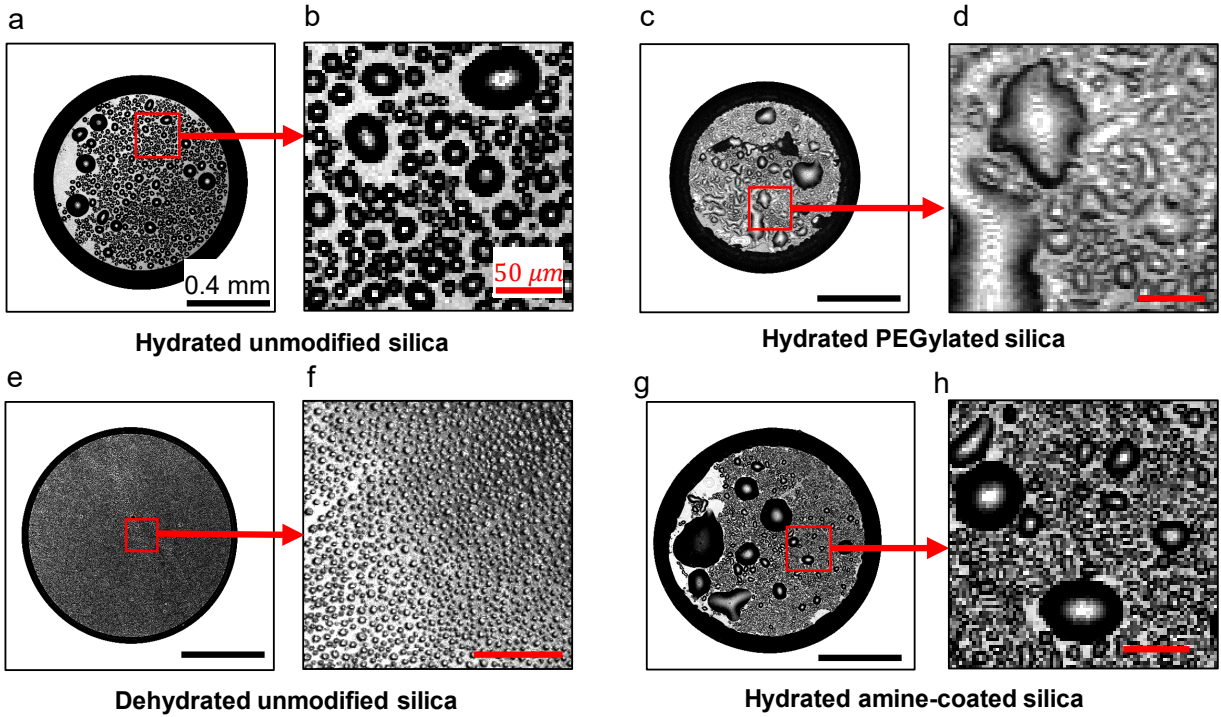

**Figure 6.** Reflection channel of confocal microscopy shows the difference in the oil microdroplets on the substrate for the different silica particles. (b), (d) and (h) are the zoomed in images of the square marked regions in (a), (c) and (g). For the case of dehydrated unmodified silica particles, (f) shows an image taken at a higher magnification of another droplet, to clearly resolve the small oil microdroplets. All images are taken at time $t/t_w \approx 0.35$, except (f) which is taken at $t/t_w \approx 0.32$. Scale bar 0.4 mm for (a), (c), (e) and (g). Scale bar 50 μm for zoomed in images. A large number of very small (diameter ~ 5 μm) oil microdroplets are nucleated on the substrate for the droplet containing dehydrated unmodified silica particles (e and f), in stark contrast to the other cases.

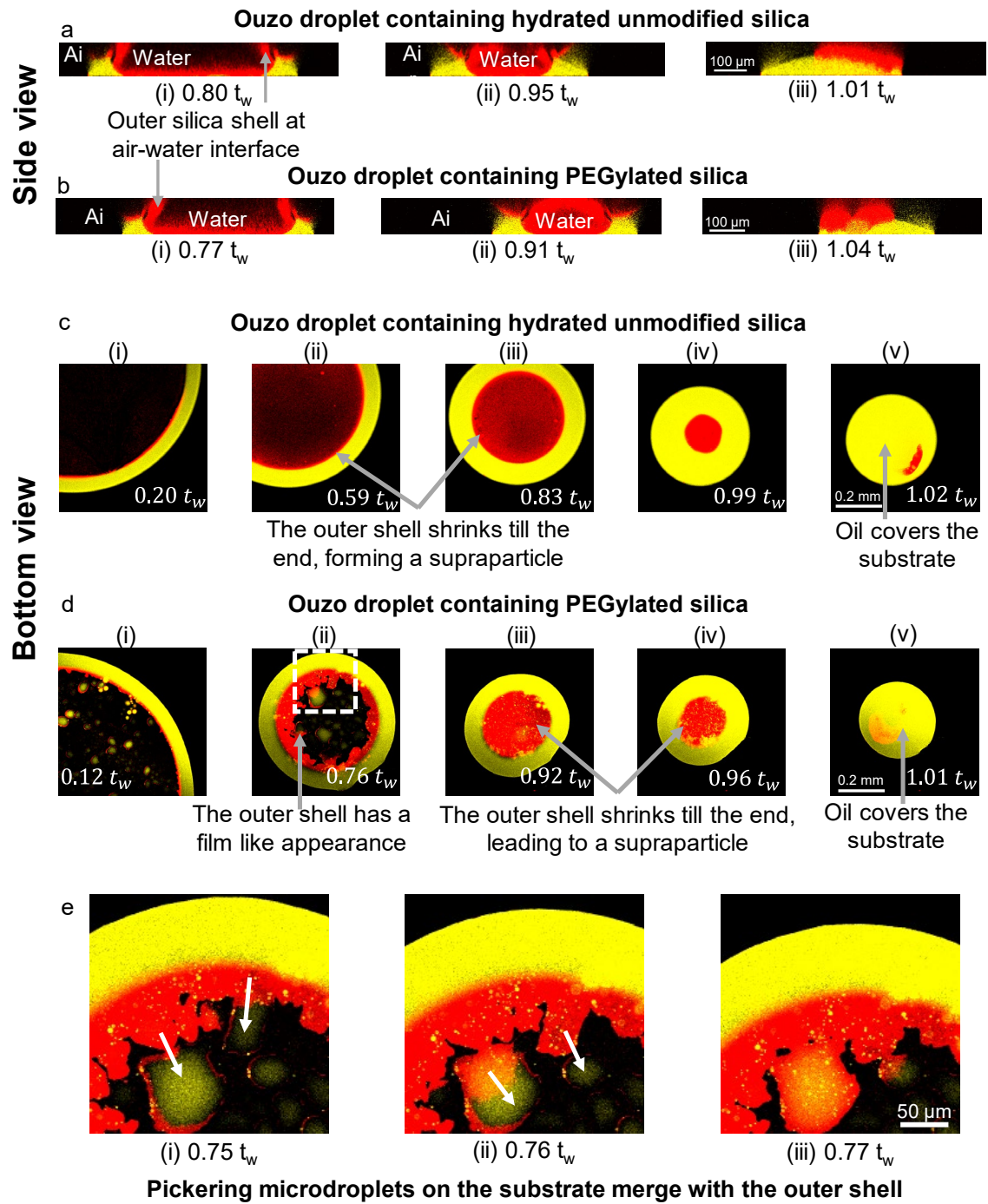

**Figure 7.** Confocal microscopy reveals similarities and differences in the formation of supraparticles in droplets containing hydrated unmodified silica (a and c) and hydrated PEG coated silica (b and d). Overlay of emission signals of perylene (yellow, showing oil) and rhodamine (red, showing silica particles) at different stages of the evaporation. (a and b) Side-view or vertical cross section, reconstructed from layer-wise scanning (Scale bar 100 µm); (c and d) bottom-view of a horizontal plane 0-10 µm from the glass substrate (Scale bar 0.2 mm). The contrast of the images has been enhanced to show all the relevant features in the images. (e) In Ouzo droplet containing PEG coated silica particles, Pickering microdroplets on the substrate merge with the outer shell, making the outer shell radially asymmetric. Images in (e) are the zoomed image of the dotted

**rectangular region in (d)-(ii). Scale bar 50 μm. The number at the bottom of each image shows the time in relation to the time $t_w$ when almost all water has evaporated, but oil is still surrounding the silica deposit. Both hydrated unmodified silica and PEG coated silica form supraparticles, even though the outer shell of PEG coated silica has a film-like appearance.**

**The hydrated silica particles** accumulate close to the oil-water and the air-water interface, similar as in our earlier study[12] (Figure 7a and c; summary of the process in Figure 9a). We refer to this particle-rich region close to the interfaces as "outer shell" for simplicity, to distinguish it from the particle shell formed by adsorption of particles on the surface of Pickering microdroplets. Over the course of evaporation, this outer shell keeps on contracting, thereby adjusting its shape to the oil-water and air-water interfaces (Figure 7a: i-ii and Figure 7c: i-iv). Close to $t = t_w$, the particles agglomerate and the outer shell becomes rigid, forming a supraparticle, which floats in the oil droplet (Figure 7a: iii and Figure 7c: v).

**PEGylated silica particles** follow a partly similar route as the hydrated silica particles, forming supraparticles (Figure 7b and d; summary of the process in Figure 9b). The particles accumulate at the interfaces and form the outer shell (Figure 7b: i-ii and Figure 7d: i-ii). Close to $t = t_w$, this outer shell stops contracting, and a supraparticle forms. However, as a result of PEGylation, Pickering microdroplets can be observed in the evaporating droplet (Figure 7d: i and ii; Figure 7e), similar as detected in the static mixtures (Figure 5b and f). The Pickering microdroplets on the substrate merge with the outer shell as evaporation proceeds (Figure 7e). As a result, the outer shell of PEGylated particles at $t≈0.76 \, t_w$ is radially asymmetric, differently from the symmetric outer shell of hydrated silica particles (Figure 7e; also compare Figure 7c: ii with Figure 7d: ii and iii). This outer shell of PEGylated particles appears like a film of particles formed at the glass-water interface (Figure 7d-ii). However, in spite of these differences, the outer shell contracts until $t≈t_w$, forming a supraparticle (Figure 7d: v).

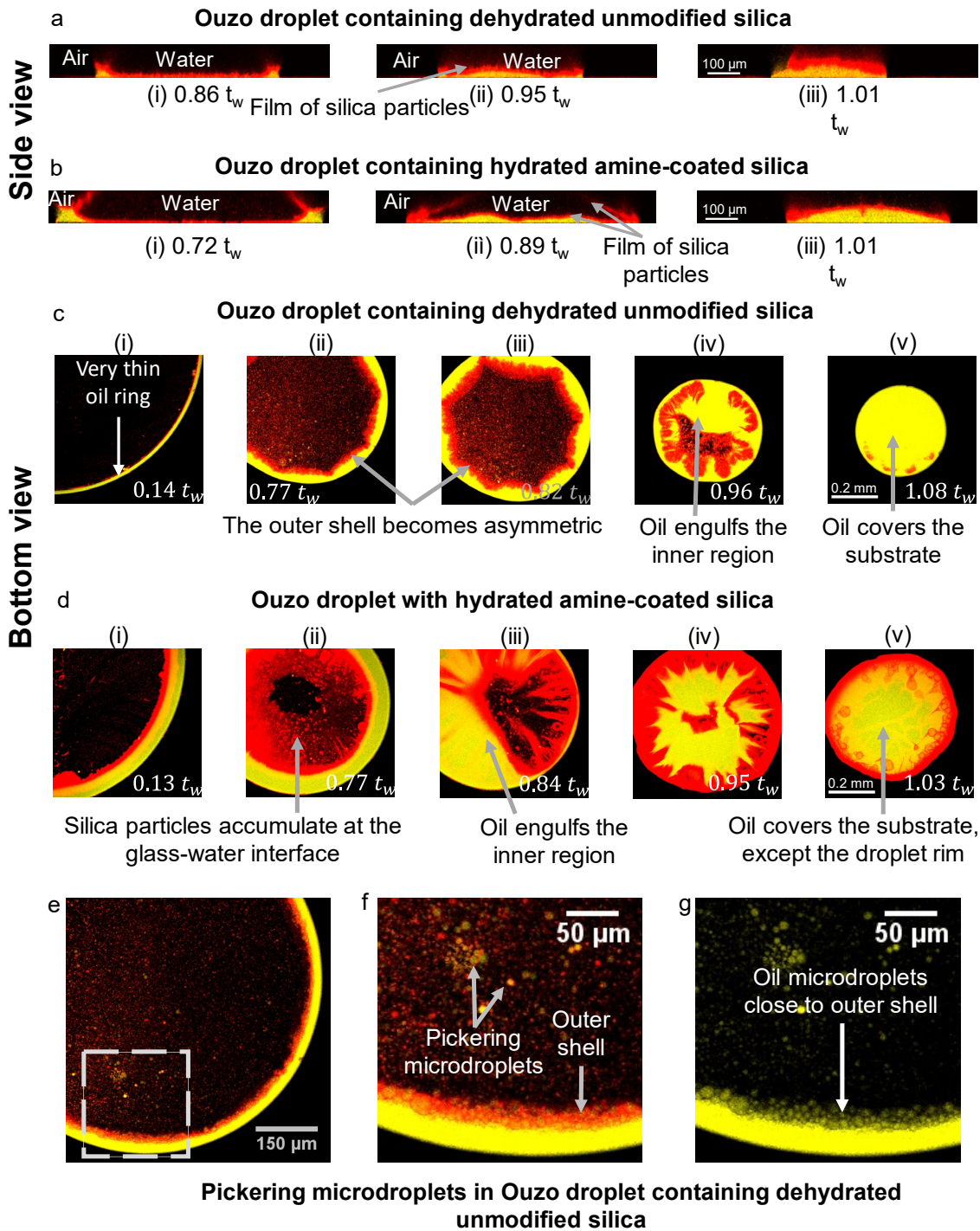

**Figure 8.** Confocal microscopy shows the formation of flat film-like structures in droplets containing dehydrated unmodified silica (a and c) and hydrated amine-coated silica (b and d). (a-d) Overlay of emission signals of perylene (yellow, showing oil) and rhodamine (red, silica particles) at different stages of the evaporation – (a and b) side-view or vertical cross section, reconstructed from layer-wise scanning (Scale bar 100 μm); (c and d) bottom-view of a horizontal plane 0-10 μm from the glass substrate (Scale bar 0.2 mm). The contrast of the images has been enhanced to show all the relevant features in the images. The number at the bottom shows the time in relation to the time $t_w$ when water has evaporated almost completely, but oil is still surrounding the silica deposit.

**(e)-(g) Pickering microdroplets observed in an evaporating Ouzo droplet containing dehydrated unmodified silica particles, at $t = 0.64t_w$. (f) and (g) are zoomed in images of the dotted rectangular region in (e). (g) Fluorescence emission signals only from perylene to show oil microdroplets close to the outer shell. Scale bar 50 µm. The evolution of outer shell at the air-water interface and close to the substrate is different for dehydrated unmodified silica particles and amine-coated silica particles. Nevertheless, both particles form flat deposits at the end of the droplet evaporation process.**

The **dehydrated unmodified silica particles** form a film like deposit (Figure 8a and c; summary of the process in Figure 9c). Notably, a large number of Pickering microdroplets is present in the water-rich central region and in the outer shell close to the oil ring (Figure 8e-g), starting early from $t \approx 0.1$-$0.2\ t_w$, similar to the static Ouzo mixtures (Figure 5c and g). Furthermore, the oil ring is much thinner, compared with all other droplets. In particular, at $t/t_w \approx 0.15$, the oil ring thickness is ≈15 µm versus ≈50 µm for all other droplets (compare Figure 8c: i with Figure 7c: i, Figure 7d: i, and Figure 8d: i). Thus, it appears that the oil that is encapsulated in the Pickering microdroplets does not contribute to the oil ring, leading to a thinner oil ring.

The particles and the oil-microdroplets accumulate at the oil-water interface, thereby forming the outer-shell similar to all the other particles (Figure 8c-i). However, the emission intensity of the dehydrated particles at the air-water interface is almost negligible in the vertical cross-section, compared to all the other particles at the same imaging settings (compare vertical cross-sections Figure 8a with Figure 7a, b, and Figure 8b). This observation suggests that the dehydrated silica particles mainly accumulate close to the oil-water and the glass-water interfaces, while comparatively less particles are present at the air-water interface (as seen in Figure 8a and Figure 8c).

As evaporation proceeds, the outer shell becomes asymmetric and non-circular around $t = 0.77\ t_w$, appearing different from hydrated and PEGylated silica particles (Figure 8c-ii). Further, around $t \approx 0.95\ t_w$, the sides of the outer shell merge together becoming a film-like structure (see SI Figure S15). At the same time, the central region gets engulfed by oil (Figure 8a-ii and Figure 8c-iv). Finally, close to $t = t_w$, the outer shell solidifies and the film-like structure of silica particles detaches from the substrate and floats in the oil (Figure 8c: v, Figure 8a: iii, and Supplementary

Video V2). This detachment of the film corresponds to the peculiar decrease in diameter and increase in height of the droplet at $t/t_w$ = 0.98 ± 0.01 in side-view shadowgraph imaging (Figure 4b: zoomed-in; Figure S6c). Finally, all the oil evaporates and a flat deposit of silica is left on the substrate.

**Amine-coated particles** follow a similar route as dehydrated unmodified silica particles and form flat film-like structures (Figure 8b and d; summary of the process in Figure 9d). However, qualitatively fewer Pickering microdroplets were detected with amine-coated particles. Thus, more oil is available to form a thicker oil ring compared to the droplets with dehydrated unmodified silica particles (compare oil ring thickness in Figure 8d: i versus in Figure 8c: i). Furthermore, there are differences in the evolution of the outer shell between the amine-coated and the dehydrated unmodified silica particles:

- Amine-coated particles form an outer shell at both interfaces — the oil-water and the air-water interfaces (Figure 8b: i and Figure 8d: i); for dehydrated unmodified particles, the outer shell at air-water interface is negligible.
- Amine-coated particles accumulate close to the glass-water interface starting around $t \approx 0.7$-$0.8\ t_w$ (Figure 8d: ii). This accumulation coincides with the pinning of the oil-glass-air contact line in base length, at $t/t_w \approx 0.7$ (shown by shadowgraphy in Figure 4b). Subsequently, the droplet undergoes consecutive depinning and pinning events until $t = t_w$, (Figure 4b). Such a pinning-depinning cycle is not seen in droplets with dehydrated unmodified silica particles.
- The conversion of the outer shell at the oil-water interface into a film-like structure and the filling of the central region by oil takes place earlier for amine-coated particles, between $t \approx 0.8\ t_w$ and $t \approx 0.9\ t_w$, compared to $t \approx 0.95\ t_w$ for unmodified dried particles (Figure 8b: ii, Figure 8d: ii,iii, and Figure S15).
- The outer shell at the air-water interface also appears film-like (Figure 8b-ii, Figure S16). Between $t = 0.9\ t_w$ and $t = t_w$, this film-like structure at the air-water interface merges with the other film-like region formed at the glass-water interface (Supplementary videos V2 and V3).

- At $t = t_w$, the entire rim of this film is in contact with the glass and only the central region is floating in the oil (Figure 8d: v, Supplementary videos V2 and V3).

Overall, the different surface properties of the particles affect their accumulation at the confining interfaces and consequently affect the motion of these interfaces. These processes determine whether or not a supraparticle or a flat film-like deposit is formed. In the following, we discuss the various mechanisms and physical principles that govern the accumulation of particles at the various interfaces.

**Particle accumulation at the air-water, oil-water, and glass water-interfaces**

Particles and oil microdroplets accumulate at the interfaces of an evaporating droplet because of the fast movement of the interfaces compared to the diffusion of particles and oil microdroplets away from the interface.[51] The competition between these two process can be characterized through a Peclet number Pe, defined as

$$Pe = \frac{t_{mass\,diffusion}}{t_{interface\,movement}},$$

where $t_{mass\,diffusion}$ is the time scale for diffusion and $t_{interface\,movement}$ is the time scale for the movement of the droplet interface. The Peclet number of silica particles of 450 nm diameter is $Pe_{silica} \approx O(100) >> 1$ (see SI Section S7 for details), and for the Pickering microdroplets $Pe_{oil} \approx O(10^3)$-$O(10^4)$ (see SI Section S7 for details). In addition, the flow field of the droplet can transport the dispersed components close to the interface.[52] As a result of these effects, the particles and Pickering microdroplets can accumulate close to the oil-water and the air-water interfaces, forming the outer shell.[12, 51, 53, 54] This shell formation was detected by confocal microscopy for all particles, except for the dehydrated unmodified silica particles at the air-water interface (compare Figure 8a with Figure 7a,b and Figure 8b).

For the dehydrated unmodified silica particles, the low concentration at the air-water interface could be because a huge amount of particles are used-up in the formation of a large number of Pickering microdroplets (Figure 8e-g). These Pickering microdroplet might sediment faster than they could be captured by the moving droplet interface. The sedimentation velocity of the Pickering microdroplets, $v_{sediment}$, can be estimated by determining the terminal Stokes velocity (Refer to SI Section 7 for details). The average velocity of the air-water interface, $v_{interface}$, can be determined from the height versus time curves of the droplet, obtained from shadowgraph measurements (Figure S6-S8). We estimate $v_{interface} / v_{sediment} \approx O(0.1) - O(0.01)$ (Refer to SI Section 7 for details). Thus, the microdroplets can sediment before getting captured by the interface, as indicated by a higher sedimentation velocity compared to the velocity of the interface. Note that the individual particles sediment much slower ($v_{interface} / v_{sediment} \approx O(1) - O(10)$; see SI Section 7 for details). Thus, the formation of a large number of Pickering droplets and their faster sedimentation rate could eventually lead to lower concentration of particles at the air-water interface. These Pickering microdroplets also have a very low Stokes number ($\approx O(10^{-7})$), suggesting that the flow field influences their motion (see SI section S7 for details). The flow field inside an evaporating Ouzo droplet is not yet completely understood, due to the hindrance in performing velocimetry, as a result of the nucleation and growth of oil-droplets.[46] We expect that a synergetic effect of the flow field and the sedimentation reduces the concentration of the dehydrated unmodified silica particles at the air-water interface and enhances the concentration preferentially at the oil-water interface where the deposit is formed.

In the case of amine-coated particles, the accumulation close to the substrate could occur due to the additional electrostatic interactions. In general, hydrophobic glass surfaces in contact with an aqueous medium show negative charge at neutral and alkaline pH values.[55-59] Therefore, we expect that the hydrophobized glass surfaces used in this study carry a negative surface charge. Thus, electrostatic interactions between positively-charged particles and negatively charged substrate can lead to accumulation of amine-coated particles close to the substrate (Figure 8b: ii).[14]

Overall, modifications in the particles' surface properties lead to differences in particle accumulations at the various interfaces. These differences eventually determine whether flat-deposits or spheroidal supraparticles are formed.

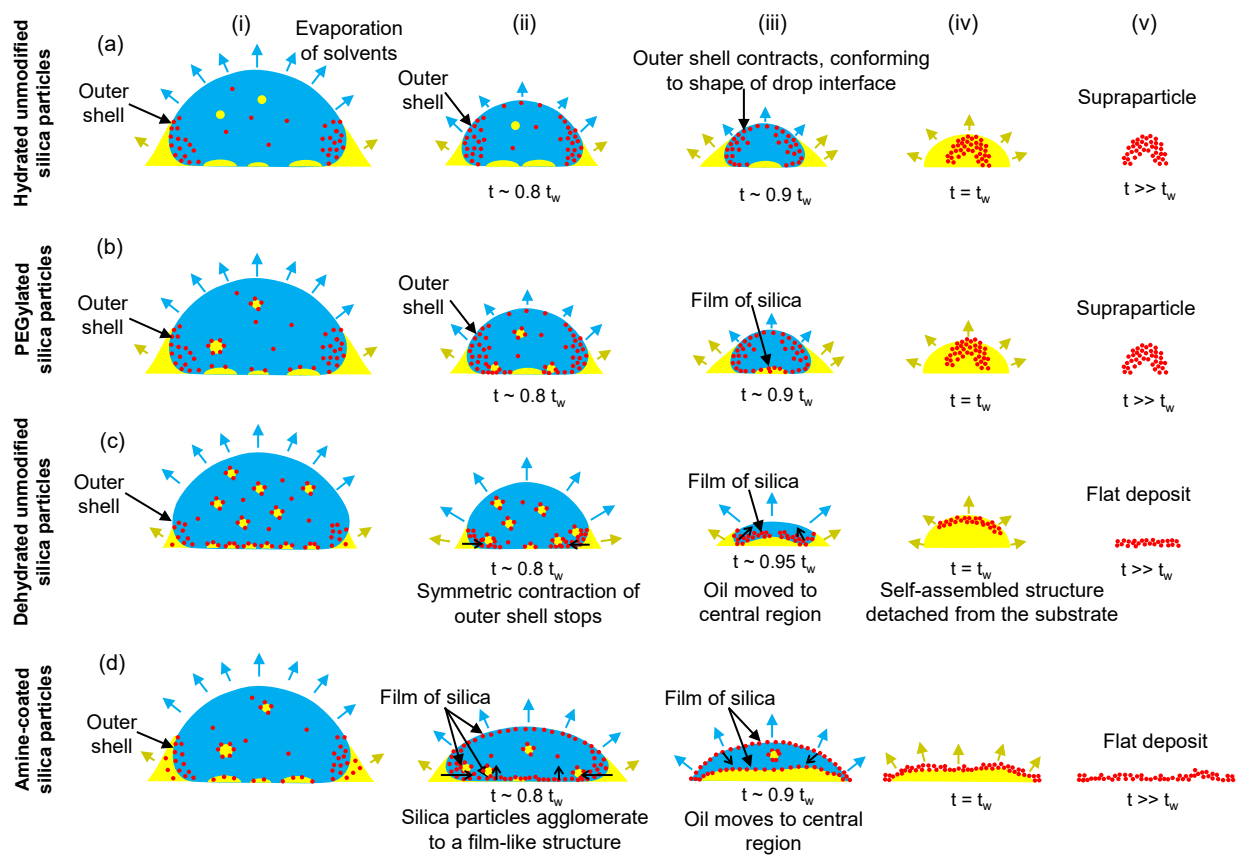

**Figure 9.** Schematic showing the mechanism of final deposits formation with the different silica particles. (a) Hydrated unmodified silica particles and (b) PEGylated silica particles form supraparticles while (c) dehydrated unmodified silica and (d) amine-coated silica particles form flat deposits. The differences in the shape of the outer shell leads to differences in the shape of the final deposit.

**Synopsis of the mechanism: flat-deposits versus supraparticles**

During the evaporation of a colloidal Ouzo droplet, high Peclet numbers lead to accumulation of particles at the moving air/water and oil/water interfaces, forming the outer shell. However, hydrated unmodified silica particles mainly did not show any detectable interactions with oil droplets or the substrate. Therefore, the outer shell can keep on contracting until close to the end of the evaporation (Figure 9a: i-iv). The shape of the outer shell is templated by the shape of the droplet interfaces. As a result, spheroidal supraparticles are formed upon the solidification of the outer shell (Figure 9a: v).

The PEGylation of particles leads to adsorption of the particles onto the oil droplets which lie on the substrate (Figure 7e). This adsorption leads to an asymmetric outer shell which appears like a film (Figure 7d). However, this film-like structure keeps on contracting (Figure 9b: i-iv). In droplets containing PEGylated particles, the steric stabilization alters the agglomeration process; in this case, the agglomeration occurs via the entanglement of polymer chains.[60] This difference in the agglomeration process and the absence of an attractive particle-substrate interaction (as both are negatively charged) could have ensured the contraction of the film and the eventual formation of supraparticle (Figure 9b: v).

In the case of unmodified dehydrated particles, the increased hydrophobicity leads to the formation of Pickering microdroplets, and an increased concentration of particles close to the substrate (Figure 9c: i-iii). Consequently, there is no shell at the air-water interface, and a flat film-like deposit forms close to the substrate (Figure 9c: iii-v).

With amine-coated particles, the contact line of the droplet undergoes severe pinning. As a result, amine-coated particles form a flat film-like deposit close to the substrate (Figure 9d: iii-v). This enhanced pinning could be due to the interactions between the positive surface-charge of the particles and the negative surface charge of the glass substrate (Figure 9d: i-iii).

Thus, the modifications in the particles' surface properties alters the behavior of the particles at the different interfaces, leading to the formation of different final deposits. Nevertheless, the interaction of colloidal particles with the various interface is governed by multiple competing effects such as Van der Waals, electrostatic (including image-charge effects), steric, capillary, and solvation interactions.[61-64] Further studies can address the role of these interactions in more detail.

Finally we compared the internal structure of the resulting deposits. The supraparticles made using self-lubricating droplets are known to have high porosity.[11] Confocal microscopy confirms the porous structure of the supraparticles made of hydrated unmodified silica particles and PEGylated silica particles, as well as of the flat deposits made of amine-coated silica particles (Figure S18 a-c, Figure S19, and Figure S20). In contrast, flat deposits made of the dehydrated unmodified silica particles did not show appreciable porous structure (Figure S18 d-f). Scanning electron microscopy (SEM) images also show differences in packing of the particles at the surface of the various deposits (see SI Section S9 for details). Thus, modifying the surface of colloidal particles affects the internal structure and the particle ordering at the surface of the final deposits. Further studies can focus on quantifying such differences in the structure and porosity of these deposits and explore ways to tune the internal and surface structure of the deposits, for instance by varying the particle to oil ratio[11] or the pH value of the dispersion.[13]

## Conclusion

In this work, we have shown the effect of changing the surface characteristics of colloidal particles on the formation of supraparticles in evaporating ternary self-lubricating Ouzo droplets. We compared the behavior of silica particles that are electrostatically stabilized — negatively charged or positively charged by amine-coating — and sterically-stabilized by PEGylation. Moreover, we studied the role of hydration of the particles. We show that these surface modifications alter the contact line motion and the spatial distribution of particles in the droplet. Consequently, particle surface modifications alter the shape of the final particle assembly. Unmodified hydrated particles accumulate as an "outer shell" close to the air-water interface and the oil ring. During the evaporation, this outer shell remains flexible, keeps on contracting, and forms a supraparticle. All the studied particle surface modifications, namely the increased hydrophobicity in dehydrated particles, the coating with positively charged amine groups, and the PEGylation, favor the interaction of colloidal particles with the oil. Consequently, in the case of sterically-stabilized particles, the outer shell at the oil-water interface additionally forms a film-like structure due to adsorption of particles on the oil droplets that are sitting on the substrate. However, the outer shell still remains flexible and a supraparticle is formed at the end of evaporation. Differently, the dehydrated negatively charged particles preferentially accumulate close to the substrate and form a flat film-like deposit. This preferential accumulation at the bottom could be because of the faster sedimentation of Pickering microdroplets compared to the accumulation of the particles at the moving air-water interface. In the Ouzo droplet containing amine-coated particles, particle-substrate interactions hinder the contraction of the contact-line, forming flat-films. These findings show the importance of carefully choosing the surface properties of colloidal particles and will enable tailoring multifunctional supraparticles for various applications such as catalysis, drug delivery, and optoelectronics, using colloidal Ouzo droplets.

## Materials and methods

### Materials

All chemicals were used a received. Tetraethoxysilane, rhodamine b isothiocyanate, *trans*-anethole (99%), trichloro(octadecyl)silane, chloroform (≥99%) from sigma-aldrich/Merck, (3-Aminopropyl)triethoxysilane (APTES) from TCI chemicals N-(6-aminohexyl)aminoproplytrimethoxysilane, (AHAPS) 92%, and 3-[methoxy(polyethylenoxy)propyl]trimethoxysilane (PEG-T) 90%, 6-9 repeat units from ABCR GmbH, Germany. For evaporation experiments, ethanol (100% absolute analytical reagent, ACS, ISO) was obtained from Boom B. V. Milli-Q water was obtained from a Reference A+ system (Merck Millipore) at 18.2 MΩ cm at 25 °C.

### Synthesis

*Rhodamine B-APTES* conjugate was synthesized similar to literature procedure.[44] Rhodamine B isothiocyanate (10 mg, 0.018 mmol) and APTES (40 μL, 37.8 mg, 0.17 mmol) were dissolved in 3 mL of absolute ethanol and stirred overnight under Ar. The resulting solution was stored at 4 °C and used without additional purification.

The synthesis of non-modified silica was done by a modified Stöber method with addition of rhodamine B-APTES conjugate.[65] TEOS (7 mL) was premixed with rhodamine B-APTES conjugate (151 µL) and added to a solution of ammonium hydroxide (16 mL, 28%) in absolute ethanol (200 mL) and ultrapure water (11 mL). After stirring over night at room temperature particles were isolated by centrifugation at 16000 g for 10 min. The pellet was washed three times with ethanol at the centrifuge and resuspended in an ultrasonic bath between the centrifugation step. After the final centrifugation step the particles were either dried overnight at 65 °C in vacuum. A fraction of particles was redispersed in absolute ethanol and kept in ethanolic solution for further modification

Surface modification of nanoparticles with PEG-T or AHAPS was done using a modified procedure from the literature.[44]

*For PEGylation*, the ethanolic dispersion of silica particles (0.201 g of particles) was diluted to a concentration of 8 g L$^{-1}$. Ammonium hydroxide (25%, 5.3 mL) was added to the ethanolic dispersions and the particles were stirred under argon for 30 min prior the addition of PEG-T. Afterwards PEG-T (0.4 mL) was added to particles, followed by refluxing overnight (80 °C) under Ar. After refluxing overnight, additional PEG-T (0.4 mL) was added to complete the modification. The reflux step was repeated one more time. Afterwards, the particles were isolated by centrifugation (15 min, 16000 g), washed three times with ethanol and dried overnight at 65 °C under vacuum.

For *modification with AHAPS*, the ethanolic dispersion of particles was diluted in the same way as for PEGylation, followed by the addition of 4.9 mL of aqueous ammonium hydroxide solution (25%, 4.9 mL). After stirring for 30 min, AHAPS (80 µL) was added, and the reaction mixtures were stirred overnight under Ar at a room temperature. Afterwards, the reaction mixtures were heated under reflux for 2 h at 80 °C. Zeta potential of a test sample was measured in-between to monitor the modification (particles isolated by centrifugation). To complete the modification, AHAPS (0.3 mL) was added to the reaction mixture. The stirring overnight and the heating step were repeated followed by the isolation of nanoparticles by centrifugation (15 min, 16000 g). The particles were purified by centrifugation with ethanol as (3 times, 15 min, 16000 g). After the final centrifugation step the pellet was dried overnight at 65 °C under vacuum.

**Characterization of silica particles.**

*DLS* was done with a Malvern Zetasizer Nano series S90 at a scattering angle of 90°. The nanoparticles were diluted with a solvent so that the attenuator was at step 9-11.

*The zeta potential* was measured with a Malvern Zetasizer Z nano using disposable cuvettes. The samples were diluted with 3 mM KCl. Zeta potentials were calculated by Malvern software from electrophoretic mobility using the Smoluchowski equation. The reported values represent a mean of five independent measurement and the standard deviation.

*Transmission electron microscopy* (TEM) was done with a JEOL1400 microscope at an acceleration voltage 120 kV. The samples were deposited on carbon-coated copper grids. The analysis of particle sizes was done with measure particles plug in for Fiji.[66]

**Confocal microscopy of the non-evaporating Ouzo mixtures.**

*Preparation of Ouzo-mixtures for confocal microscopy in dispersion*.

To prepare the Ouzo-mixtures with dried particles, silica particles were redispersed in ethanol using an ultrasonic bath at a concentration of 10 mg mL$^{-1}$. The stock solution was used to prepare a mixture of particles and *t*-anethole in ethanol. To label the oil, 3 μL of perylene-solution in ethanol were added. Subsequently, water was added quickly to induce the phase separation. The vial was closed quickly and vortexed for 1-3 s. The imaging was started 10-30 min after addition of water.

The samples with hydrated particles were prepared as following: silica particles redispersed in water at a concentration of 10 mg mL$^{-1}$ and incubated in water over night. Subsequently, the dispersions were diluted with water to the respective concentration. In a different vial, a solution of *t*-anethole in ethanol was prepared and labeled by addition of perylene. Aqueous dispersion of particles was quickly added to the oil to induce the Ouzo-effect or the phase separation followed by vortexing for 1-3 s. The imaging was started 15-30 min after preparation of the samples.

The weight fractions of components in the Ouzo-mixture were the following: *t*-anethole / particles / water / ethanol: 0.020 / 0.0016 / 0.55 / 0.43.

*Confocal microscopy* (Leica TCS SP5, Wetzlar, Germany) was used to assess the droplet size distribution and adsorption of silica particles onto the droplets. Grace Bio-Lab SecureSeal imaging spacers were used to form a well on a coverslip that was filled with approximately 50 µL of the Ouzo mixture. Subsequently the well was covered with a second coverslip to avoid the evaporation. Images were recorded with a 63X NA1.2 water immersion objective, oil phase (labeled with perylene) was excited at 458 nm and emission was detected with a HyD in the spectral range from 473.5 to 528.8 nm. For particles (rhodamine B), excitation was 561 nm, and emission (597-761 nm) was recorded using a photomultiplier. For all images excitation intensities and gain were identical to allow for a quantitative comparison of the Ouzo mixtures. TileScans consisting of 3x3 images were acquired, representing an imaged area of approximately 650x650 µm (See supplementary data Z1 for full size images). Due to the flow, distortions of the droplets can occur in the overlap region of the automatically merged images. Position and size of the droplets were evaluated using the DetectCircles plugin of ImageJ. Shell intensities (thickness) were estimated from the particle fluorescence channel, plotting 36 lines with tilt angle increasing by 10°, extending from the center of the droplet to 1.3 times the radius of the droplet. The maxima of the intensities along these lines in the region between 0.4 and 1.3 times the radius were averaged to calculate the shell intensity for one droplet. These intensities were subsequently averaged over all droplets.

**Preparation of Ouzo mixture for evaporation experiments**

To prepare the colloidal Ouzo mixtures with "dehydrated particles" (Figure 1) for evaporation experiments, the silica particles were put in a glass vial and then ethanol, *t*-anethole, and water were added to the vial, in this order. The mixture was sonicated for at least 5 min after adding each liquid.

To prepare the colloidal Ouzo mixtures with hydrated particles (Figure 1) for evaporation experiments, the silica particles were put in a glass vial and then water was added. The mixture was sonicated for at least 5 min and then incubated overnight. On the next day, ethanol and *t*-anethole were added to the mixture. The mixture was sonicated for at least 5 min before adding each liquid. The weight fractions of the colloidal Ouzo mixture were as follows: *t*-anethole / particles / water / ethanol — 0.014/0.001/0.453/0.532

**Preparation of substrates (glass slides) for evaporation experiments**

The glass substrates were hydrophobized by dip-coating, by modifying the procedure mentioned in the literature.[11, 12] The glass slides are initially cleaned by mechanical wiping, sonication in solvents, and plasma-treatment. Meanwhile, some water saturated solution of chloroform and toluene (in 2:1 ratio, by volume) is prepared (hereby called water saturated solvent). Thereafter, a solution of chloroform and toluene, in the ratio of 1:4, by volume, is prepared (hereby called main solvent). The water saturated solvent is added to the main solvent, such that volume of the water-saturated solvent is 2% of the volume of the main solvent. Finally, OTS is added to this mixture, such that the volume of OTS is 0.45% of the initial main solvent. The OTS-solution is mixed properly. Immediately thereafter, the cleaned glass slides are put inside the OTS-solution for hydrophobization. After 30 min, the glass slides are put in a solution of chloroform and sonicated for 10 min, to remove all the unreacted OTS from the glass surface. After 10 min, the glass slides can also be rubbed by a chloroform wetted cotton swab, if any whitish marks of non-bound OTS are visible. The glass slides are then rinsed with ethanol and water to remove all chloroform from the glass. Finally, the glass slides are blow-dried with nitrogen and stored until they are used. The substrates had an advancing contact angle of 113° ± 3° and a receding contact angle of 106° ± 6°.

**Experimental Setup for shadowgraph**

Details of experimental set-up can be found in Raju *et al.*.[12] In short, the evaporating droplets were observed from the side using monochrome 8-bit CCD camera and from the top using a CMOS color camera. Both the cameras were connected to a Navitar 12X adjustable zoom lens. LED light sources were used for illumination. A thermo-hygrometer (OMEGA; HHUSD-RP1) was used to measure the ambient temperature and humidity. The temperature was measured as 21 ± 1 °C and the relative humidity was measured as 60 ± 10%.

**Characterization the final deposits**

The SEM of the final deposits was done with a Zeiss MERLIN HR-SEM.

**Confocal microscopy of evaporating droplets**

For performing confocal microscopy on evaporating droplets, Nikon Confocal Microscopes A1 system 667 (with 10× and 20× dry objectives) were used. The bottom view images in Figure 7 and Figure 8 were obtained using the Galvano mode, while the vertical cross-sections were obtained by reconstructing data from successive horizontal planes, captured in resonant mode. To collect the images from all the horizontal planes and create a single snapshot of a vertical cross-section, it takes ≈ 4 s.

**Image processing of evaporating droplets**

The side-view shadowgraph images were analyzed using MATLAB-R2020 and FIJI.[66] Details of this image processing can be found in Raju *et al.*.[12]

**Acknowledgement**

The authors thank E. Muth (MPIP), C. Sieber (MPIP), and H. Therien-Aubin (MPIP) for measurements and discussions, and S. Schumacher for the preparation of supplementary cover art. We also thank M. Smithers from the NanoLab at University of Twente for the help with SEM. We acknowledge the funding from Max Planck University of Twente Center for Complex Fluid Dynamics. D.L. also acknowledges the funding from ERC Advanced Grant DDD (no. 740479). X.H.Z. acknowledges the support by the Natural Sciences and Engineering Research Council of Canada (NSERC) and Future Energy Systems (Canada First Research Excellence Fund) and the funding from the Canada Research Chairs program. OK acknowledges the funding from Alexander von Humboldt Foundation.

**Supporting Information**

Supporting information contains the following

- Sections S1-S9: Adhesion energy of particles at liquid-liquid interfaces; additional images of deposits from evaporation experiments and their characterization; additional data on the evaporation process; additional characterization of the static Ouzo mixtures; fluorescence images showing the merging of outer-shell of dehydrated unmodified silica particles and the film-like structure of particles-assembly in droplets containing amine-coated silica particles; detailed analysis of sedimentation time, Stokes number, Peclet number; internal structure of the supraparticles and the flat deposits; the arrangement of the colloidal particles at the surface of the deposit (SEM images)
- Video V1: Video of top-view and side-view (shadowgraph) imaging of evaporating Ouzo droplets.
- Video V2: 2-D vertical cross-section of evaporating colloidal Ouzo droplets using fluorescence confocal microscopy.
- Video V3: 3-D reconstruction of evaporating colloidal Ouzo droplets using fluorescence confocal microscopy.
- Zip file Z1: Full-size images of oil and particle distribution in non-evaporating Ouzo mixtures observed using confocal microscopy. Smaller cropped regions from the images in Z1 are used in Figure 5 and Figure S9.

**Present address** OK: Sustainable Polymer Chemistry Lab, Department of Molecules and Materials, Faculty of Science and Technology, University of Twente, PO Box 217, 7500 AE, Enschede, The Netherlands

**Graphical Abstract**

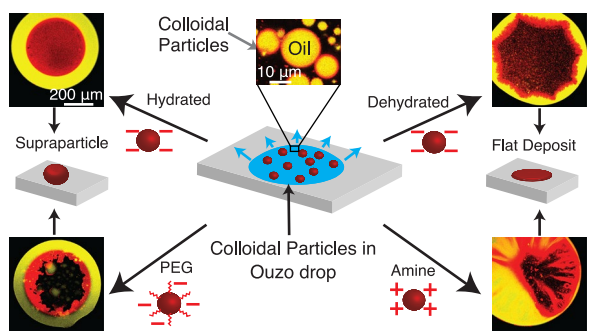